\documentclass[12pt]{article}
\usepackage{amsmath,amsfonts,graphicx,yfonts,mathabx,braket}
\usepackage[nosort]{cite}
\newcommand{\be}{\begin{equation}}
\newcommand{\ee}{\end{equation}}
\newcommand{\bea}{\begin{eqnarray}}
\newcommand{\eea}{\end{eqnarray}}

\newcommand{\mt}[1]{\textrm{\tiny #1}}
\newcommand{\Tr}{{\rm tr}}

\newcommand{\la}{\lambda}
\renewcommand{\title}[1]{\vbox{\center\LARGE{#1}}\vspace{3mm}}
\renewcommand{\author}[1]{\vbox{\center#1}\vspace{3mm}}
\newcommand{\address}[1]{\vbox{\center\em#1}}
\newcommand{\email}[1]{\vbox{\center\tt#1}\vspace{3mm}}

\begin{document}
\begin{titlepage}
\begin{center}
\rightline{\tt}
\vskip 2.5cm
\title{{\bf Replica trick and string winding}}
\vskip .6cm
\author{Andrea Prudenziati and Diego Trancanelli}
\vskip -.5cm 
\address{Institute of Physics, University of S\~ao Paulo\\ 05314-970 S\~ao Paulo, Brazil}
\vskip -.1cm 
\email{prude, dtrancan@if.usp.br}
\end{center}
\vskip 3cm

\abstract{
\noindent  
We apply the replica trick to compute the entropy of a cylinder amplitude in string theory. We focus on the contribution from non-perturbative winding modes and impose tadpole cancellation to understand the correct prescription for integrating over moduli. Choosing the entangling surface to cut longitudinally over the whole length of the cylinder, we obtain an answer that is interpreted as the entropy of a density matrix. We recast this result in target space language, both in the open and closed string picture.
}
\vfill
\end{titlepage}


\section{Introduction}

The replica trick is a powerful method to compute path integrals associated to density matrices. It is employed in the most diverse fields, from statistical physics and machine learning to conformal field theory, especially in relation with the study of entanglement, see {\it e.g.} \cite{wilczek,replica,dot,machine,glass}. 
In general, given a region $A$ at some fixed time in a manifold $\mathcal{C}$, the replica trick allows to compute the $q$-th power of the reduced density matrix $\rho_A$. This is the partial trace $\rho_A=\Tr_{B} \rho$ of the full quantum state $\rho$ defined at fixed time in ${\cal C}$ over the degrees of freedom contained in the complement of $A$, called $B$ here. Explicitly, one has \cite{wilczek,replica}
\begin{equation}\label{rt}
\Tr(\rho_A^q)=\frac{Z_q(A)}{(Z_1(A))^q}\,,
\end{equation} 
where $Z_q(A)$ is the path integral over a $q$-sheeted surface obtained by gluing together $q$ replicas of the original manifold $\mathcal{C}$ after having performed a cut along $A$, which is called entangling surface. The upper lip of the cut on one sheet is identified with the lower lip of the cut in the next sheet. Equipped with this trace, the associated von Neumann entropy can be computed by analytically continuing $q$ to a real number, differentiating with respect to it, and finally setting $q=1$.

\begin{figure}[ht]
\centering
\vspace{-0pt}
\includegraphics[width=0.8\textwidth]{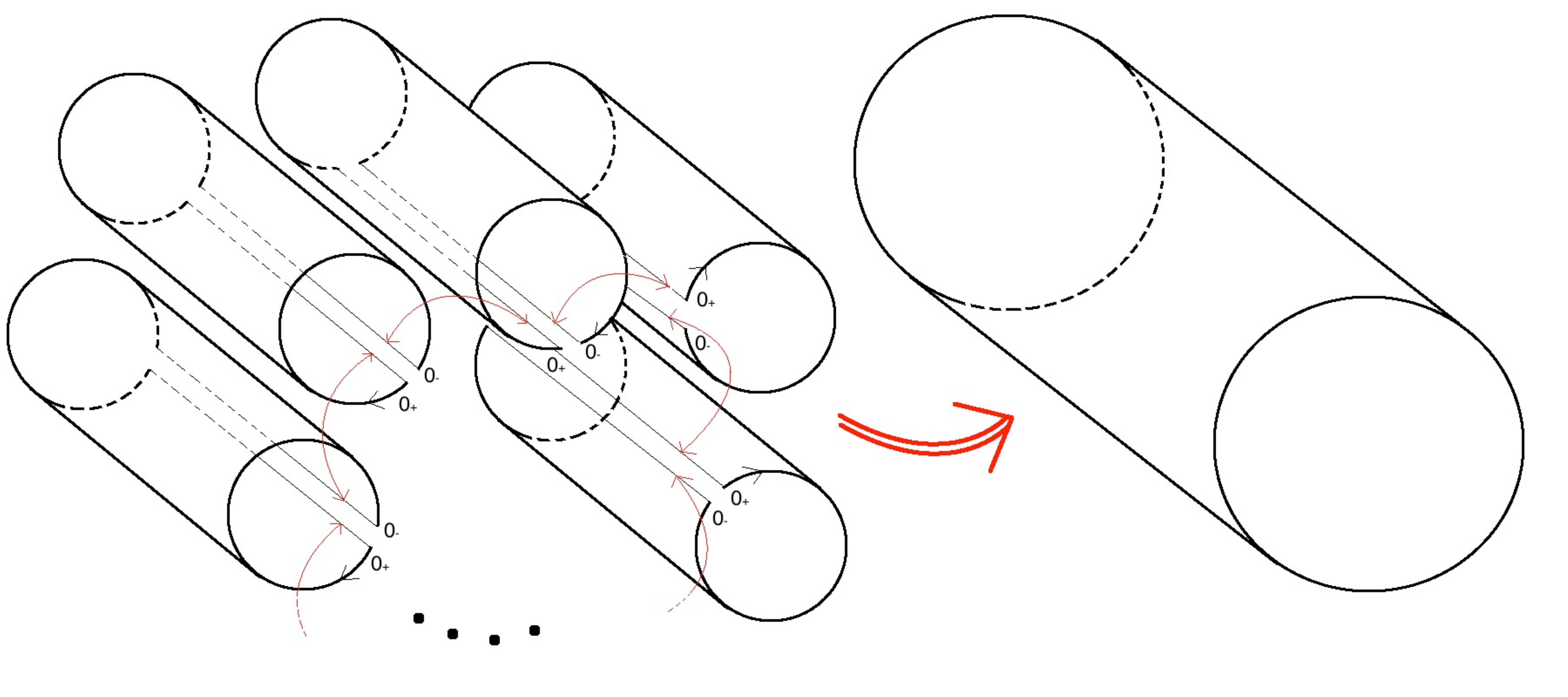}
\vspace{-0pt}
\caption{$q$ cylinders are cut longitudinally, excising an infinitesimally thin strip between world-sheet times $0_-$ and $0_+$, and then glued together. The cut at $0_+$ on the $i$-th sheet is glued to the cut at $0_-$ on the $(i+1)$-th sheet. Finally, the cut at $0_+$ on the $q$-th sheet is identified with the cut at $0_-$ on the first sheet, resulting in a large cylinder. }
\label{bigcyl}
\end{figure}

The goal of this note is to apply the replica trick to a basic string theory object: the cylinder amplitude. We want to compute (\ref{rt}) for the case in which $\mathcal{C}$ is the cylindrical world-sheet formed by a propagating string and $A$ is taken to cut longitudinally over the whole length of the cylinder, from one boundary to other. The resulting surface is depicted on Fig.~\ref{bigcyl}.
The world-sheet time in the figure parametrizes the compact dimension. It is obvious that, in this case, $Z_1(A)$ is the usual cylinder amplitude, which shall be called $Z$ in what follows.
Note that, as the entangling surface at fixed world-sheet time covers the full world-sheet, we are not tracing over any external degrees of freedom. In other words, $B$ is empty in this setup. The output of the computation is then the usual von Neumann entropy -- rather than some entanglement entropy -- of whatever quantum state, either pure or mixed, has been produced by the path integral at that time. The result for the entropy is non-vanishing if and only if that state is a density matrix.

The above statement, relating the usual thermal entropy with the entanglement entropy in the limit of a vanishing complement of the entangling region, comes from a holographic intuition \cite{Azeyanagi:2007bj}. 
For a system of size $L$, one has
\be\label{therm}
\lim_{\epsilon\to 0} \left(S_\mt{ent}(L-\epsilon)-S_\mt{ent}(\epsilon)\right)=S_\mt{therm}\,.
\ee
 This can be interpreted as the difference in length between a geodesic revolving around the dual black hole horizon and a vanishing one,\footnote{More precisely, this happens when the entangling region size approaches the UV cut-off and finite terms depend on the cut-off definition.} so that the entropy is really produced by the first term alone. From the replica trick point of view, this translates into the physical requirement that $\lim_{\epsilon\to 0}Z_q(\epsilon)= (Z_1(\epsilon))^q$ (times some power of $\epsilon$).

Strings have both localized, oscillatory modes carrying momentum and extended, topologically inequivalent configurations characterized by winding numbers. Here we shall focus on the non-perturbative contribution originating from the latter modes, for the case in which the target space is either a two-dimensional torus or a circle.
We do not include the contribution from the string oscillations. This is interesting for a variety of reasons. First of all, it is one of the first attempts\footnote{See also the recent \cite{Donnelly:2016jet} for a computation of entanglement entropy in the string theory dual of 2-dimensional Yang-Mills theory. More in general, some references which dealt with entanglement entropy in string theory include \cite{PandoZayas:2014wsa,He:2014gva,Das:2015oha,Hartnoll:2015fca,Schnitzer:2015gpa,Zayas:2016drv,Vancea:2016tkt}.} to use the replica trick on the string world-sheet, which is tricky because of the obvious topological nature of the theory. Here this construction is well defined as the cut stretches from one boundary of the world-sheet to the other, so it is itself ingrained in the topology. Second, we will consider non-perturbative contributions that, to the best of our knowledge, have only been studied before, for entanglement entropy, in \cite{non-pert}. Third, we will deal with some subtleties in the definition of moduli on the world-sheet replicas by requiring tadpoles cancellation, displaying yet another application of this mechanism. Finally, the result will be expressed in terms of target space quantities, like the complex structure and the open string moduli (positions of the branes), thus establishing a precise map between the entropy of the cylinder and the geometrical properties of the embedding space. This map will be further checked by deriving the world-sheet results from a target space perspective, both in the open and closed string channels.

  
\section{The replica trick on the world-sheet}

We start with the world-sheet computation of the entropy. To this scope, we consider open strings stretching between parallel D-branes. As stated above, we only study the case in which the world-sheet has the topology of a cylinder and the entangling surface is taken to cover its whole longitudinal length.

The target space we consider is a 2-dimensional torus, times non-compact spectating dimensions that we suppress for not playing any role in the analysis. Our notation is the following. The torus has two cycles parameterized by the vectors $R_1$ and $R_2 e^{i\alpha}$, with $R_1, R_2, \alpha\in \mathbb{R}$, and complex structure $\sigma=\frac{R_2 }{R_1}e^{i\alpha}$. In terms of target space coordinates $x^a=(x^1, x^2)$, related to the standard complex coordinates $(z,\bar z)$ by $z=R_1 x^1 +  R_2 e^{i\alpha}x^2$, the Wilson line for the Chan-Paton factor on the $i$-th brane of the stack is $A_i=\theta_i dx^1-\phi _i dx^2$. 

There are a number of equivalent ways to do the computation, which are related by T-duality. We shall consider D0-branes at positions determined by the Wilson lines. These are obtained by T-dualizing twice the space-filling D2-branes with $A_i$, as follows. In terms of the coordinates $x^a$, the metric is found to be
\begin{equation}\label{metricx}
g_{11}
=R_1^2\,,\qquad 
g_{12}
=R_1 R_2 \cos{\alpha}\,,\qquad
g_{22}
=R_2^2,
\end{equation}
with the complex structure $\sigma$ and the torus area $t_2=R_1 R_2 \sin{\alpha}$ being given by
\begin{equation}\label{cstgij}
\sigma=\frac{g_{12}}{g_{11}}+i \frac{\sqrt{\det(g)}}{g_{11}}\,, \qquad t_2=\sqrt{\det(g)}\,.
\end{equation}
The introduction of a B-field $b=b_{12}\, dx^1\wedge dx^2$,  with $b_{12}= R_1 R_2 \cos{\alpha}$, complexifies the area: $t=t_1+i t_2=b_{12}+i \sqrt{\det(g)}$. Doing  a T-duality along $x^1$ exchanges area and complex structure. We see then that $R_1\rightarrow \frac{1}{R_1}$ exchanges $\sigma$ and $t$. Similarly, a T-duality along $x^2$ transforms $t\rightarrow -\frac{1}{\sigma}\sim \sigma$ and $\sigma \sim -\frac{1}{\sigma} \rightarrow 
t$ (the sign $\sim$ means that $\sigma$ and $-\frac{1}{\sigma}$ define the same complex structure). The sequence of these two T-dualities maps (\ref{metricx}) to the dual metric
\be\label{dm}
\tilde{g}_{11} = \frac{1}{R_1^2} \,,
\qquad \tilde{g}_{12} =-\frac{\cos{\alpha}}{R_1 R_2} \,,\qquad
 \tilde{g}_{22} =\frac{1}{R_2^2}\,,\qquad 
 \tilde{b}_{12}= -\frac{\cos{\alpha}}{R_1 R_2}\,.
\ee
The D2-brane with Wilson line $A_i$ is mapped into a D0-brane at position $\theta_i$ along $x^1$ and $-\phi_i$ along $x^2$. An open string with Chan-Paton factors $i,j$ at the endpoints extends between two D0-branes localized at $(\theta_i,-\phi_i)$ and $(\theta_j,-\phi_j)$, respectively. 

The cylinder world-sheet $\mathcal{C}$ can be parameterized by a coordinate $\ell_1\in[0,1]$, stretching between the two D0-branes, and a coordinate $\ell_2\in[0,2\pi s)$, which is attached on the D0-branes at the endpoints $\ell_1=0,1$. As a consequence, there are two independent winding numbers for the string when moving along $\ell_1$: the string can in fact wrap $m$-times along $R_1$ and $n$-times along $R_2 e^{i\alpha}$. Note that, due to the cyclic identification of the sheets along the entangling surface, one can only have one pair of winding numbers $(n, m)$ common to all sheets. The classical maps $X^a$ from the world-sheet $(\ell_1,\ell_2)$ to the target space are
\begin{equation}
\label{maps}
X^1=\ell_1(m+\theta_{ij})\,,\qquad 
X^2=\ell_1(n-\phi_{ij})\,,
\end{equation}
with $\theta_{ij}\equiv \theta_i-\theta_j$ and $\phi_{ij}\equiv \phi_i-\phi_j$.
The corresponding path integral weight $e^{-S_\mt{cl}}$ can be obtained by evaluating\footnote{Note that the contribution from the B-field is zero, as it necessarily contracts with $\partial_{\ell_2}X^{1,2}=0$.}
\bea
S_\mt{cl}=\int d^2\ell \;\partial_{\ell_1} X^a\partial_{\ell_1} X^b \tilde{g}_{ab}
=\sin^2\alpha\frac{2 \pi s}{t_2 \sigma_2} \left| n+\sigma m- u_{ij}\right|^2\,,\qquad
u_{ij}\equiv \phi_{ij}+\sigma \theta_{ij}\,.
\label{acr}
\eea

Computing the winding number contribution to the amplitude $Z$ for $\mathcal{C}$ is quite simple. The prescription is to integrate (\ref{acr}) over the modulus $s$ and to sum over the winding numbers:
\be
\label{cylpre}
Z=\sum_{m,n}\int_0^{\infty}\frac{ds}{4s}\exp\left(-\sin^2\alpha\frac{2 \pi s}{t_2 \sigma_2} |n+\sigma m- u_{ij}|^2\right) \,.
\ee
Evaluating the integral and the sums is a well-known procedure, see for example \cite{Bonelli:2009aw}. A Poisson resummation in $n$ and $m$ needs to be performed. Discarding the $n=m=0$ term, one can sum the series and obtain the amplitude
\be
\label{cyl}
Z=-\frac{1}{2}\log\left| e^{\pi i \theta_{ij}^2\sigma}\theta_1(u_{ij}|\sigma)\eta(\sigma)^{-1}\right|\,,
\ee
in terms of the standard modular functions $\theta_1$ and $\eta$, see for example \cite{singer}. 

A couple of comments are now in order. First, the mechanism of tadpole cancellation plays an important role here. Given an appropriate choice of gauge group, it cancels the divergent $n=m=0$ terms between the cylinder, the M\"{o}bius strip and the Klein bottle, leaving a finite series for non-vanishing Wilson lines \cite{Bonelli:2009aw}. Second, the $n=m=0$ modes would be the only ones depending on the prefactors in the action~(\ref{acr}), but this dependence disappears from (\ref{cyl}). 

The amplitude $Z_q(A)$  comes from considering $q$ cylinders, cutting them longitudinally and gluing the cuts together, thus obtaining a big cylinder out of $q$ rectangles, as shown in Fig.~\ref{bigcyl}. To compute this, one needs to understand the correct prescription for integrating over moduli. At first sight, there appear to be two possibilities, which depend on the order in which the operations of cutting/gluing cylinders and integrating/modding out the world-sheet metric by the diff$\times$Weyl symmetries are carried out.

Suppose one first integrates and mods out by diff$\times$Weyl on each copy of the cylinder and then cuts and glues them together. The first operation yields, as usual, $q$ integrals over moduli from diff$\times$Weyl-inequivalent metrics, one for each cylinder. When cutting and gluing the cylinders, the fields on adjacent sheets should be identified across the cuts, so that the moduli also get identified, leaving a single surviving modulus. 
Analogously, the measure factors of $1/4s$ appearing in each integral from translation invariance along the cylinder are reduced to a single one, as translation along a sheet moves the cut and leads to the same translation along all other cuts. In the action (\ref{acr}), instead of a single multiplicative factor of $s$, we now have a factor of $qs$, from the sum of $q$ actions with the same modulus and winding numbers. This first integration prescription leads to
\be
\label{ch1}
Z_q(A)=\sum_{m,n}\int_0^{\infty}\frac{ds}{4s}\exp\left(-\sin^2\alpha\frac{2 \pi q s}{t_2 \sigma_2} |n+\sigma m-u_{ij}|^2\right) \,.
\ee
The other possible choice is to first cut and glue the $q$ cylinders and then to integrate over diff$\times$Weyl-inequivalent metrics. In this case, one first constructs the big cylinder and then deals with the path integral over the world-sheet metric, leading to a single modulus measuring the entire circumference of the cylinder. This second option gives just the usual cylinder (\ref{cylpre}).
We note at this point that the factor of $q$ that appears in the exponent in (\ref{ch1}) can be simply reabsorbed by a rescaling of the prefactor. As seen above, the final result will be independent of this prefactor after tadpole cancellation has been applied. In summary, the two procedures are equivalent and the two operations of metric integration and cutting/gluing do in fact commute.\footnote{A similar issue appears when dealing with fermions. In that case the question is whether the sum over spin structures has to be done on each sheet before or after the replica. The two procedures lead to different results, see for example \cite{Azeyanagi:2007bj,Zayas:2016drv,Herzog:2013py,Lokhande:2015zma}, unlike what happens in the present case.}

The Renyi entropy (\ref{rt}) is at this point easily obtained\footnote{As $\rho_A=\rho$, we suppress from now on the somewhat misleading label $A$, which may induce to believe we are computing entanglement entropy, when in fact we are computing the usual entropy.}
\bea
\label{re}
\Tr\left(\rho^q\right)
&=&\left(-\frac{1}{2}\log\left| e^{\pi i \theta_{ij}^2\sigma}\theta_1(u_{ij}|\sigma)\eta(\sigma)^{-1}\right|\right)^{1-q}\,.
\eea
The entropy of $\rho$ is given by differentiating this expression with respect to $q$ and setting in the end $q=1$, which results in
\be
S^\mt{torus}=-\Tr\left(\rho \log \rho \right)=-\partial_q \Tr(\rho^q)|_{q=1}=\log\left(-\frac{1}{2}\log\left| e^{\pi i \theta_{ij}^2\sigma}\theta_1(u_{ij}|\sigma)\eta(\sigma)^{-1}\right|\right)\,.\, 
\label{ee}
\ee
This quantity $S^\mt{torus}$ measures the non-perturbative contribution to the entropy of the quantum state $\rho$ propagating on the cylinder with the cut $A$. This result would be trivially zero if $\rho$ were pure. We can therefore conclude that this state is a density matrix. We shall discuss the nature of this density matrix below, when we recast it in a target space language in terms of open and closed string channels. 

The reader may also note at this point that we have not introduced any cut-off in the result. Indeed, it is well-known that the replica trick is essentially blind to cut-offs
(which can, however, be seen as coming from curvature singularities at the branch points \cite{replica}). In the present case, a cut-off will have to be introduced below, from requiring that the entropy be real. 

Let us now specialize to the case when the target space is not a torus, but a circle of radius $R$. This will be useful in the following. We can obtain the cylinder amplitude $Z$ from (\ref{cylpre}) by considering a purely imaginary complex structure $\alpha = \pi/2$, resulting in $\sigma=i\sigma_2=i\frac{R_2}{R_1}$ and $t_2=R_1R_2$, a Wilson line $-\phi_i$ along the $x^2$ direction, and by removing the winding number $m$. After an obvious renaming of $R_2$, the amplitude reads
\begin{equation}\label{cyls1}
Z=\sum_{n}\int_0^{\infty}\frac{ds}{4s}\exp\left(-\frac{2 \pi s}{R^2} (n-\phi_{ij})^2\right)\,,
\end{equation}
whose evaluation is straightforward. One first Poisson resums and then does the integral, with the $n=0$ term removed as before, obtaining
\bea
\label{cyls2f}
Z&=&
\sum_{n\neq 0}\frac{1}{4 |n|}e^{2\pi i n \phi_{ij} }
=-\frac{1}{2}\log|1-e^{2\pi i\phi_{ij}} |\,.
\eea
The entropy of $\rho$ on the target space circle becomes
\begin{equation}\label{ees1}
S^\mt{circle}=\log\left(-\frac{1}{2}\log|1-e^{2\pi i\phi_{ij}} |\right)\,.
\end{equation}
Note that the result is not defined for values of $\phi_{ij}$ such that (\ref{cyls2f}) is negative, namely $1/6<\phi_{ij}<5/6$. To avoid this, we introduce by hand a cut-off. Not having dimensionful quantities to guide our choice and being the replica trick itself blind to cut-off dependence, we can just guess a viable {\it ad hoc} answer which is replacing (\ref{ees1}) with
\begin{equation}\label{ees1c}
S^\mt{circle}=\log\left(-\frac{1}{2}\log\left(\mu |1-e^{2\pi i\phi_{ij}} |\right)\right)\,,
\end{equation}
for $1/2\ge \mu\ge 0$. The resulting entropy as a function of the distance $\phi_{ij}$ is symmetric around $\phi_{ij}=1/2$,  monotonically decreasing in the first half, and convex. 


\section{Target space derivation of the density matrix}

\begin{figure}[ht]
\centering
\vspace{-0pt}
\includegraphics[width=0.7\textwidth]{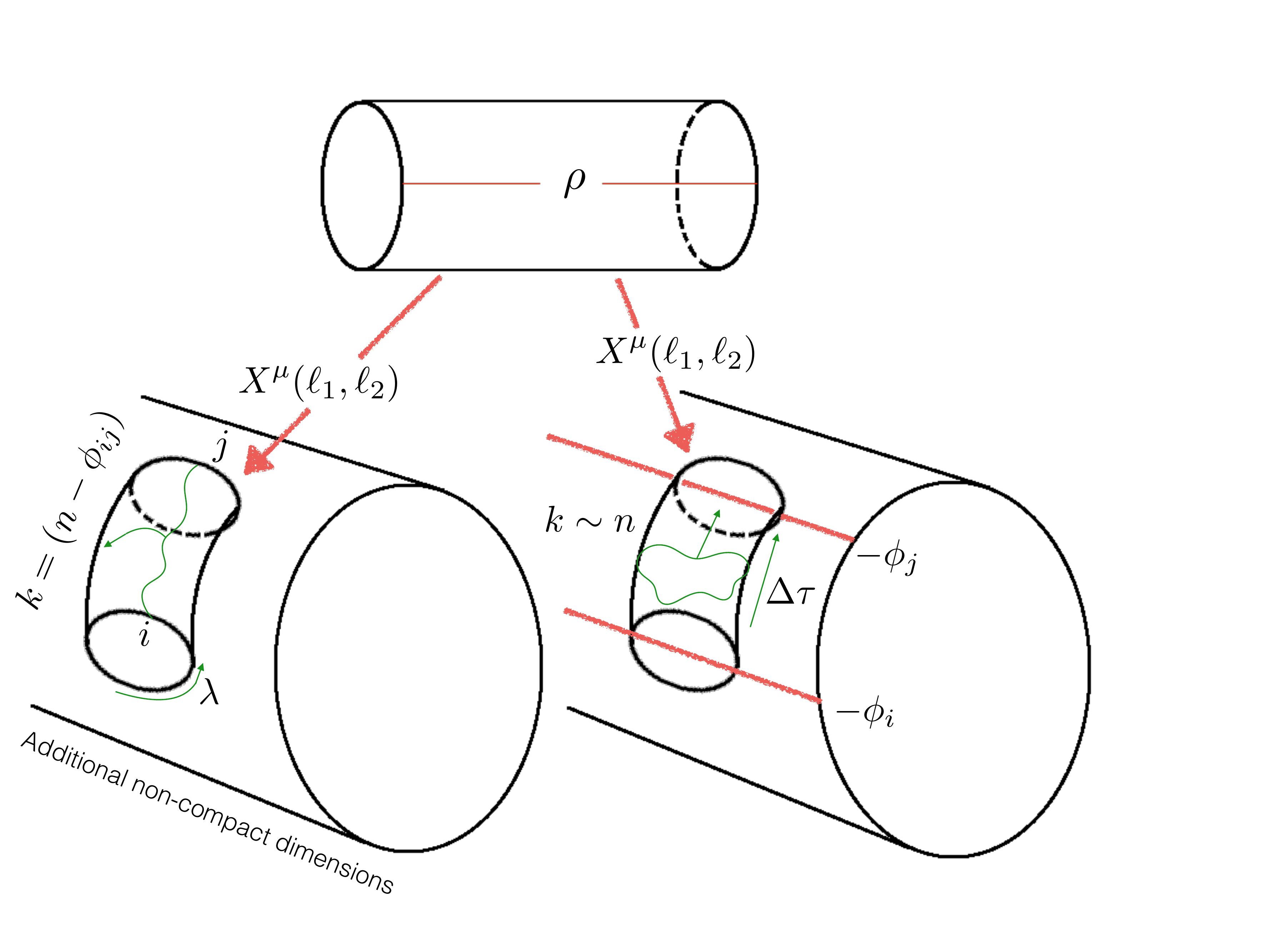}
\vspace{-0pt}
\caption{
Target space interpretation of the world-sheet density matrix $\rho$, in the open (left) and closed (right) string channels. The open string stretches between two D1-branes and carries momentum $k=n-\phi_{ij}$, whereas the closed string stretches between two D0-branes and carries momentum $k\sim n$.
}
\label{target}
\end{figure}

We want to interpret these results from a target space point of view, using a construction for open and closed strings summarized in Fig.~\ref{target}. The main idea is to select an appropriate target space density matrix which reproduces the world-sheet expressions. 

\subsection{Open string picture}

We initially look at open strings, corresponding to space-filling D-branes with Wilson lines. Limiting our attention to the center of mass of the strings and suppressing their oscillations, as done above in (\ref{maps}), we can think of them as particles charged under the Wilson lines. To further simplify things, we consider a circle instead of a torus target space, thus aiming to reproduce (\ref{ees1}). The density matrix we consider is a sum over projectors, each corresponding to a particle of momentum $k$, charged under the $i,j$  Wilson lines, 
and that has propagated along a path of length $2\pi\lambda$, see Fig.~\ref{target} (left). The path integral weight for such a particle is given by $\frac{1}{4\lambda}\exp(-2\pi\lambda k^2)$ (see for example \cite{Polchinski:1998rq}) and we get
\begin{equation}
\label{tss}
\rho=\frac{1}{\Tr \rho_u}\rho_u\,, \qquad
\rho_u=\sum_{n}\int_0^{\infty}\frac{d\la}{4\la}\,e^{-2\pi\lambda k^2}\ket{n,\la}\bra{n,\la}\,.
\end{equation}
  As the integration measure from modding out translations, time reversal invariance and the $i\leftrightarrow j$ symmetry of unoriented strings is $\frac{d\la}{4\la}$, we change variables to $\gamma(\lambda)=\frac{1}{4}\log\lambda$ so that the measure in (\ref{tss}) becomes $d\gamma$. We emphasize that we are integrating over the particle world-line length, 
rather than the modulus of the cylinder world-sheet (to which it reduces, however, when the string length goes to zero). The states $\ket{n,\gamma}$  are taken to be orthonormal in both discrete and continuous labels: $\braket{n,\gamma|n',\gamma'}=\delta_{nn'}\delta(\gamma-\gamma')$. The momentum $k$ is quantized on the circle and shifted by the Wilson line
\be\label{mom}
k^2=\frac{1}{R^2}\left(n-\phi_{ij}\right)^2\,.
\ee
To compute the entropy we first construct $\rho^q$,
whose trace is\footnote{Formally, this should be obtained by considering first a generic matrix element of $\rho^q$ and integrating over all states as $\int d\gamma\int d\gamma' \braket{\gamma|\rho^q|\gamma'} $. The trace should be then seen as coming from one of the two states, say $\ket{\gamma'}$, approaching the other, {\it i.e.} restricting one of the two domains of integration to a vanishing interval $O_\gamma$ centered around the value of $\gamma$ of the other integral: $\Tr(\rho^q)=\lim_{O_\gamma \to 0}\int d\gamma\int_{O_\gamma} d\gamma' \braket{\gamma|\rho^q|\gamma'} $. Evaluating the delta functions before the limiting process gets rid of 
factors of $\delta(0)
$ 
that would otherwise appear in (\ref{rhoak2}).}
\begin{equation}
\label{rhoak2}
\Tr(\rho^q)=\frac{1}{(\Tr \rho_u)^q}\sum_{n}\int_{-\infty}^{\infty} d\gamma\;e^{-\frac{2\pi q \la(\gamma)}{R^2}\left(n-\phi_{ij}\right)^2}\,.
\end{equation}
 The normalization $(\Tr \rho_u)^q$ is similarly evaluated
\be
(\Tr\rho_u)^q=\left(\sum_{n}\int_{-\infty}^{\infty} d\gamma\;e^{-\frac{2\pi \la(\gamma)}{R^2}\left(n-\phi_{ij}\right)^2}\right)^q\,.
\ee
 Remembering that the $q$ in the exponential of (\ref{rhoak2}) can be rescaled away thanks to tadpole cancellation, one finds that in terms of the original $\lambda$ variable 
\be
\Tr(\rho^q)=\left(
\sum_{n}\int_0^{\infty}\frac{d\la}{4\la}\,e^{-\frac{2\pi  \la}{R^2}\left(n-\phi_{ij}\right)^2}
\right)^{1-q}\,,
\ee
which leads precisely to the world-sheet expression (\ref{ees1}) for $S^\mt{circle}$.


\subsection{Closed string picture}

A complementary approach to recover $S^\mt{circle}$ from the target space is to consider closed strings. It is well-known that an open string cylinder between two D-branes can be thought of as a closed string propagation, see Fig.~\ref{target} (right). The target space circle will be in the T-dual picture with respect to the previous section, where branes are positioned on their Wilson line values $-\phi_i$ and $-\phi_j$. The density matrix $\rho$ will be a linear combination of projectors corresponding to all possible propagations of a closed string between the $i$-th and $j$-th brane, with all possible momenta (depending on an integer $n$) and winding numbers $m$ around the circle. The weights will be expressed in the boundary state formalism:
\be\label{state}
\rho_u=
\frac{\pi R^2}{2}\sum_{n,m}\int_{\substack{ \\ \\ \\  X(\tau_i)=\phi_i \;\;\;\;\;\;\\ X(\tau_f)=\phi_j +m}} \hspace{-1.8cm} dx \, d(\Delta \tau) \ket{n,m;x,\Delta \tau} \bra{n,m;x,\Delta \tau}\braket{B,\tau_f|B,\tau_i}\braket{B,\tau_f|B,\tau_i}^{\dagger}\,.
\ee
Here $x$ and $\Delta\tau\equiv \tau_f-\tau_i$ represent, respectively, the position at $\tau=0$ and the time of propagation for the classical part of the euclidean string map:
$
X(\tau)=x-i 
\tau k 
,
$
in units of $\alpha'$. The quantity $\braket{B,\tau_f|B,\tau_i}$ is the scalar product between the two boundary states for the closed string cylinder, representing the two D-branes at path times $\tau_i$ and $\tau_f$ (for the closed string world-sheet, this is the longitudinal coordinate). We also have that 
\begin{equation}\label{evo}
\ket{B,\tau_i}=e^{(\tau_i-\tau_f)(L_0+\tilde{L}_0)}\ket{B,\tau_f}\,,
\end{equation}
and $\braket{B,\tau|B,\tau}=1$. Orthonormality in all indices is assumed. We can then repeat the procedure of the previous section. The only difference is that the weights in  (\ref{rhoak2}) will be replaced by the ones in (\ref{state}). 

The first step in the evaluation of this amplitude is to enforce the conditions on the embedding $X(\tau)$ to start and finish on the proper brane, after having wound up around the circle $m$-times. These conditions are delta functions that will fix the initial position of the map $x$ and $\Delta\tau$ (being the origin for $\tau$ irrelevant by translational symmetry). The condition at $\tau_i$ translates into a delta $\delta(X(\tau_i)-\phi_i) $ that becomes just the condition $x=\phi_i+i  \tau_i k$. The condition at $\tau_f$ becomes instead $\delta(X(\tau_f)-\phi_j-m) $ which can be transformed into a delta for $\Delta\tau$ 
\begin{equation}\label{dt}
\delta(X(\tau_f)-\phi_j-m)
=\delta\left(\Delta \tau +i\frac{\phi_{ij}-m}{
k}\right)\frac{1}{| 
k|}.
\end{equation}
We can now exploit (\ref{evo}) with $L_0=\tilde{L}_0=k^2/ 4 R^2$. The momentum $k$ is quantized around the circle,
$k=2\pi R^2 n$ for some integer $n$. 
Putting everything together, we obtain (the dagger also exchanges $i$ and $j$)
\be
\braket{B,\tau_f|B,\tau_i}
=
e^{i\pi n (\phi_{ij}-m)}
\,,
\qquad
\braket{B,\tau_f|B,\tau_i}^{\dagger}=
e^{i\pi n (\phi_{ij}+m)}
\,.
\ee
Combining this with (\ref{dt}) one obtains $e^{2\pi i n \phi_{ij}}/2
\pi R^2 |n|$, having the dependence on $m$ disappeared. 
After removing, as usual, the $n=0$ term from the sum, the final result for the Renyi entropy reads
\begin{equation}\label{cc}
\Tr(\rho_A^q)=\left(\sum_{n\neq 0} \frac{1}{4|n|}e^{2\pi i n \phi_{ij}}\right)^{1-q}\,,
\end{equation}
whose corresponding entropy is in perfect agreement with (\ref{ees1}).


\section{Discussion}

We have computed the entropy associated to the winding modes of a string wrapping a torus, finding a non-vanishing result that depends only on the complex structure $\sigma$ of the torus and on the position of the branes on which the string endpoints are attached. It is worth emphasizing that this entropy does not originate from localized degrees of freedom, but from extended, topologically inequivalent configurations. This is the first time, to the best of our knowledge, that the replica trick has been applied to non-perturbative states and for a string world-sheet. We have chosen a prototypical problem to clarify important issues. The first one is about the correct moduli prescription on the $q$-sheeted cylinder, which has been solved by applying the tadpole cancellation mechanism to show equivalence between two different approaches. It would be interesting to understand if this idea can be generalized to more complicated Riemann surface topologies. Moreover, although the result we have obtained is non-singular, a cut-off has to be introduced in order to guarantee reality and positivity of the entropy. The replica trick is insensitive to the cut-off and our proposal in (\ref{ees1c}) is quite {\it ad hoc} and most likely not unique. Exploring the origin and justification for the cut-off in (\ref{ees1c}) is certainly something important to look at with more care, which is something we leave for the future. Finally, we have reproduced the result from a density matrix constructed using target space quantities, both in the open and closed string channels, making transparent the origin of the entropy we have found. 

It would also be interesting to obtain the full result for the entropy, including perturbative contributions, as the natural expectation of the entropy being the one associated with a canonical ensemble, with inverse temperature given by the cylinder radius, is spoiled by the necessity to integrate over this radius, {\it i.e.} the modulus of the cylinder. However, in order to do this computation, some non-trivial issues with the replica trick on the cylinder need to be sorted out first. These include: possible boundary contributions \cite{Berthiere:2016ott}; the fact that string theory is conformal only locally, so that the usual CFT computations should be adapted somehow;\footnote{In particular, the operations of integrating over diff$\times$Weyl-inequivalent metrics and cutting/gluing the Riemann surfaces might not necessarily commute, so that the corresponding expressions (\ref{cylpre}) and (\ref{ch1}) might not be equivalent in general.} the topological structure of the world-sheet that may create problems in defining non-topological entangling regions as, for example, a region that does not extend from one boundary to the other;\footnote{This would be the case if one were to do the computation by considering first the entanglement entropy for a finite entangling region and by successively  sending this region to cover the whole world-sheet.} the still unresolved issue on how to sum over spin structures \cite{Zayas:2016drv,Lokhande:2015zma}. This last problem, in particular, is quite relevant, as computations on a torus have shown that modular invariance is preserved only if appropriate spin structures are chosen. More specifically, one has to sum over each sheet and then glue sheets together in the limit of a small entangling region, whereas for a large entangling region one has to select the same spin structure for each sheet first and then sum over structures \cite{Lokhande:2015zma}. These two procedures were inspired by imposing (\ref{therm}). We wonder whether similar issues arise for the world-sheet moduli as well, whenever the entangling region does not span the whole length of the world-sheet. Finally, a last challenge would be the target space interpretation of this full result. 

On a more speculative level, it would be worthwhile to explore the implications of our computation to the understanding of the density of string states and the transitions related to winding modes. Perhaps, our entropy could be used as some order parameter for the Hagedorn behavior \cite{Barbon:2004dd}. We hope to come back to these explorations in the near future.


\subsection*{Acknowledgements}

We are happy to thank Rafael Chavez, Leo Pando Zayas, Tadashi Takayanagi and Daniel Teixeira for comments and feedback. DT acknowledges partial financial support from CNPq and from the FAPESP grants 2014/18634-9 and 2015/17885-0. He also thanks ICTP-SAIFR for their support through FAPESP grant 2016/01343-7.


\end{document}